\documentclass[aps,prd,twocolumn,amsmath]{revtex4}
\usepackage[dvips]{graphicx}

\begin{document}

\title{Properties of a Discrete Quantum Field Theory}

\author{Micheal S. Berger}
\email{berger@indiana.edu}
\author{Naoki Yamatsu}
\email{nyamatsu@indiana.edu}
\affiliation{Physics Department, Indiana University,
Bloomington, IN 47405, U.S.A.}

\begin{abstract}
A scalar quantum field theory defined on a discrete spatial 
coordinate is examined. The renormalization of the lattice propagator is 
discussed with an emphasis on the periodic nature of the associated 
momentum coordinate. The analytic properties of the scattering amplitudes 
indicate the development of a second branch point on which the branch cut from 
the optical theorem terminates. 
\end{abstract}


\maketitle

\section{Introduction}

Quantum mechanics can be and has been formulated on compact manifolds. The
fact that the  
momenta becomes quantized is well-known. The usual Fourier 
transform which connects the position space of the compact manifold
to momentum space
becomes a discrete Fourier transform. Topological features can arise when 
quantizing a theory 
on a manifold. Some 
of the interesting global aspects arise even in the simple case where the 
configurations space is a circle ($S^1$) and the phase space is its cotangent
bundle. See Ref.~\cite{Isham} for a review 
and discussion of the mathematical details. Since the phase space is a 
symplectic manifold, one can also interpret a quantum theory where the roles
of the configuration space and the momentum space are reversed. While the 
interpretation of the theory as the quantization of a classical system may be 
absent, the theory is nevertheless a well-defined quantum system.
One straightforward
technique for producing a well-defined quantum mechanics on 
a discrete configuration space is to consider momentum space as 
being compact, in which case it is the position space that 
becomes discrete\cite{Bang:2006va}. If one properly defines a 
Hamiltonian on this 
discrete space, one can represent various dynamics and one can 
produce in the continuum limit (where the discretization becomes
small) various continuum Hamiltonians (for example the free 
particle)\cite{Bang:2008vf}. For wave packets localized in 
momentum space the topology has only a small effect on the dynamics
and one obtains the usual time evolution in the continuum limit.   

One can also consider the case where both position space and 
momentum space are discrete. In fact this may be the more familiar 
case to most physicists. When the compact manifolds are taken to 
be (discrete) circles, then the phase operators which replace the 
usual position and momentum operators obey the Weyl
algebra\cite{Schwinger}.  
The study of the algebra is extensive in physics.

These ideas can be extended to quantum field theory. Indeed the 
idea of quantum field theory defined on a configuration 
space which is a compact manifold has
a rich history. In Kaluza-Klein theories the spacetime symmetries of 
extra dimensions can be seen as internal symmetries in the four-dimensional
theory. Compactified dimensions have played a major role in 
recent years in string 
theory and later in particle physics phenomenology where possible extra 
dimensions can be rendered unobservable at low energies by making them 
sufficiently small or by making the metric describing them highly warped. Many
studies assume that each extra dimension is topologically a circle to make 
the analysis tractable. 
Some of the physical effects of these extra dimensions can be 
accounted for by modifying the propagator to include corrections arising 
from the possibility of winding 
around the compact direction\cite{Cheng:2002iz}. 
One can also define a quantum field on discrete spacetime coordinates. This
approach, known as lattice gauge theory, can be formulated as one  
with the momentum space, as the dual to the configuration space, being
a compact space. Lattice physicists usually describe this feature as integration
over a Brillouin zone, but here we would like to emphasize the interpretation
in terms of a compactification of momentum space.
One then obtains a quantum field theory defined on 
discrete spacetime coordinates (or lattice), 
and again some of the physical effects
are represented by a modified propagator. In lattice gauge theory the 
discretization is a technical device invented to obtain approximate results
which become increasingly accurate as one approaches 
the continuum limit. 

In this paper 
we investigate a lattice propagator with one discretized dimension in 
configuration space. We investigate the renormalization of the theory, 
calculate some results at the one-loop level, and 
analyze the analytic properties of scattering 
amplitudes. 
For this purpose it is necessary and interesting to consider
momentum scales which correspond to 
distance scales much {\it smaller} than the lattice spacing. 
We consider a scalar field theory for simplicity.
Since we assume weak coupling the calculations here are most similar to those
which are performed in lattice gauge theory to understand the continuum 
limit\cite{Capitani:2002mp}.

Attempts to formulate theories of quantum gravity often 
contain the idea of a minimal length either explicitly or 
implicitly. In some models spacetime is formulated directly 
in terms of some quantum degrees of freedom whose characteristic
size is given by the Planck scale. In string theory the 
physical extent of the string means that experimentally one
cannot in principle probe length scales less than the Planck scale
in a scattering experiment. These physical requirements may serve as 
motivation for studying at theories which have an ultraviolet cutoff (say 
from some kind of discretization), but 
nevertheless admit practical calculations.

\section{Scalar Quantum Field Theory on a Discrete Configuration Space}

We will consider a real scalar field with a momentum space defined 
as $M^3\times S^1$ with coordinates $(p^0,p^1,p^2,p^3)$. 
This corresponds to defining a 
Hamiltonian on the configuration space with one dimension having an 
equal and discrete interval spacing,
\begin{align}
H&=\int\frac{dxdy}{(2\pi)^2}
\sum_n\left[
\Pi(x,y,z_n)^2
+\left(\frac{\partial\Phi(x,y,z_n)}{\partial x}\right)^2
\right.\nonumber\\
&\left.\hspace{-1em}
+\left(\frac{\partial\Phi(x,y,z_n)}{\partial y}\right)^2
+\left(\frac{\Phi(x,y,z_{n+1})-\Phi(x,y,z_n)}{z_{n+1}-z_{n}}\right)^2
\right.\nonumber\\
&\left.\hspace{-1em}
+m^2\Phi(x,y,z_n)^2+\lambda \Phi(x,y,z_n)^4\right],
\label{Ham}
\end{align}
where $\ell\equiv z_{n+1}-z_n$. The requirement of equal spacing is not 
required, but makes the calculations that follow tractable and the usual
propagator description applicable. Also unrequired is
the assumption that the discretized derivative involves only adjacent lattice
sites. The position space fields $\Phi(x,y,z_n)$ and $\Pi(x,y,z_n)$ can be
written by the momentum space fields $\varphi(p)$ and
$\pi(p)$ 
\begin{align}
\Phi(x,y,z_n)&=\int\frac{d^3p}{(2\pi)^3}
\varphi(p)e^{i(p_1x+p_2y+p_3z_n)},\\
\Pi(x,y,z_n)&=\int\frac{d^3p}{(2\pi)^3}
\pi(p)e^{i(p_1x+p_2y+p_3z_n)}.
\end{align}
Substituting these relation to the Hamiltonian, we obtain the
Hamiltonian on momentum spaces with one compactified momentum space,
\begin{align}
H&=\frac{1}{2}
\int\frac{d^3p}{(2\pi)^3}
\bigg[\pi(p)\pi(-p)\nonumber\\
&\ \ \ 
+\left\{p_1^2+p_2^2
+\frac{4}{\ell^2}\sin^2\left(\frac{\ell p_3}{2}\right)+m^2\right\}
\varphi(p)\varphi(-p)\bigg].
\end{align}
not including the interaction.
We emphasize that the $p_3$ integration is over a circle of circumference
$2\pi/\ell$. This is performed in a frame we call the preferred frame which is 
also the one in which the Hamiltonian is defined. If one boosts in the 
$3$-direction the standard Lorentz contraction will cause the separation between
the adjacent points to decrease. The preferred 
frame therefore corresponds to the
one where the points are defined to be separated by the largest distance at 
equal times. In momentum space the integration corresponds to one where the 
cylinder is parameterized such that the $p_3$ direction is defined in this 
preferred frame. 
Clearly the generalization that the momentum space manifold has
to be treated with open patches for the integration. Here we avoid this 
technical problem by staying in the preferred frame at all times during the 
calculation.

The Hamiltonian leads to the free propagator
\begin{align}
\tilde{D}(p)=
\frac{i}
{p_0^2-p_1^2-p_2^2
-\frac{4}{\ell^2}\sin^2\left(\frac{\ell p_3}{2}\right)-m^2}.
\label{propagator_S1}
\end{align}
We also make sure that this gives the ordinary propagator on four
dimensional Minkowski space for $\ell\to 0$
\begin{align}
\tilde{D}(p)=\frac{i}{p_0^2-p_1^2-p_2^2-p_3^2-m^2}
=\frac{i}{p^2-m^2}.
\end{align}
The propagator in Eq.~(\ref{propagator_S1}) is properly defined as periodic
for a $p_3$ coordinate defined on a circle.

The discretization of the spatial coordinate can also be thought of as a 
means of regularization of the ultraviolet divergences. For the case 
$M^3\times S^1$ the effect is to replace divergences in a four-dimensional 
theory with those of a three-dimensional one. So by the standard power
counting arguments, the quadratic 
divergence of the scalar self-energy is softened to a logarithmic divergence
and so on. 
Just as the regulated divergences 
cancel in physical observables in a renormalized field theory, low energy
cross sections and other measurables are independent 
of the scale $\ell$ to leading order. The $\ell$-dependent corrections are,
however, calculable, and here represent small effects of 
the physical scale of discretization.

One can of course proceed to discretize the other spatial directions and even 
the time direction as is usually done in lattice theory. 
We restrict our attention to the case of one discrete 
coordinate because it is sufficient to highlight the features we want to 
demonstrate.

The appearance of the sine function in the propagator is a consequence of 
assuming only nearest neighbor interactions in the 3-direction. 
We can consider more generally any periodic function $f(p_3^2)$ instead 
and relate the propagator to an effective Hamiltonian.
This discussion is sufficient to consider only the one discrete spatial 
dimension, and 
we can formally write 
\begin{align}
H=\int\frac{dp}{2\pi}
\left[\pi^2+f(p)\varphi^2\right].
\end{align}
We shall examine the constraints on this function $f(p)$.
First since $f(p)$ satisfies a periodic condition 
$p\to p+2\pi/\ell$, we can write
\begin{align}
f(p)=f(e^{iqp\ell}),
\end{align}
where $q$ is any integer. 
Second, $f(p)$ is real in order to satisfy the hermiticity of Hamiltonian.
\begin{align}
f(p)=f\left(\cos\left(qp\ell\right),\sin\left(qp\ell\right)\right).
\end{align}
Third, invariance under the parity transformation requires $f(p)=f(-p)$, so
that
\begin{align}
f(p)=f\left(\cos\left(qp\ell\right)\right).
\end{align}
Fourth, $f(p)$ is equal to zero for $p=0$
\begin{align}
f(p)=f\left(\sin^2\left(\frac{qp\ell}{2}\right)\right).
\end{align}
Finally, $f(p)$ is equal to $p^2$ when $\ell$ goes to zero. Then
\begin{align}
f(p)=\frac{1}{\sum_{q=1}a_q}
\sum_{q=1}\frac{4a_q}{(q\ell)^2}\sin^2\left(\frac{qp\ell}{2}\right),
\end{align}
where the $a_q$'s are real number and $1/\sum_{q=1}a_q$ is a normalization
factor needed to normalize the factor of $p^2$ equal to one when $\ell$ goes to
zero. 

We can write a general Hamiltonian satisfying the above constraints
\begin{align}
H&=\int\frac{dp}{2\pi}
\left[\pi^2
+\frac{1}{\sum_{q=1}a_q}
\sum_{q=1}\frac{4a_q}{(q\ell)^2}\sin^2\left(\frac{qp\ell}{2}\right)
\varphi^2+m^2\varphi^2
\right]\nonumber\\
&=\sum_n
\left[\Pi^2(z_n)+\frac{1}{\sum_{q=1}a_q}
\sum_{q=1} a_q\left(\frac{\Phi(z_{n+q})-\Phi(z_n)}{z_{n+q}-z_n}\right)^2
\right .
\nonumber\\
&\ \ \ \left .+m^2\Phi(z_n)^2
\right],
\end{align}
Therefore the interpretation is that the nearest neighbor coupling in 
Eq.~(\ref{Ham}) becomes a Hamiltonian which includes couplings between lattice
sites of any separation (represented here by $q\ell$).

\section{Loop Effects}

In this section we investigate, using the theory with discretized space,
the full propagator and a scattering cross section using Feynman rules derived
from Eq.~(\ref{Ham}). The lattice propagator in Eq.~(\ref{propagator_S1})
can be understood as a tree level
propagator defined for a periodic momentum space, or alternatively
it can be expanded 
in terms of an infinite series of operator insertions in the usual continuum 
propagator. These insertions can be understood as those contained in field 
theory containing Lorentz-violating coefficients\cite{Colladay:1998fq}, 
but we do not pursue this further here. We define
$-iM^2(p)$ as the sum of all 1-particle-irreducible(1PI) insertions into
the propagator. The full 2-point function is given by the geometric
series, 
\begin{align}
G(p)=\frac{i}{p^2-m^2}+(\mbox{terms regular at}\ p^2=m^2).
\label{propagator}
\end{align}
In this expression the four-momentum squared is defined on $M^3\times S^1$
as 
\begin{align}
p^2=p_0^2-p_1^2-p_2^2-\frac{4}{\ell^2}\sin^2\left(\frac{\ell p_3}{2}\right),
\label{psquare}
\end{align}
This guarantees that the propagator has the correct properties to describe an
asymptotic state. The mass-shell condition is expressed as $p^2=m^2$. Since 
Lorentz invariance is broken, an interesting consequence is that the wave 
function renormalization $Z$ of the propagator must be a function of $p_3$, so 
it is not a constant. 

We can define renormalization conditions for $M^3\times S^1$.
A subtlety that emerges is that one must specify a renormalization 
point for the momentum in the 3-direction since the Lorentz symmetry is 
broken. For $M^3\times S^1$, we adopt the renormalization conditions that the
pole in this full propagator occur at
$\tilde{p}^2=p_0^2-p_1^2-p_2^2=m^2$ and $p_3=0$ and have residue 1.
\begin{align}
\left.M^2(p)\right|_{\tilde{p}^2=m^2,p_3=0}&=0,\\
\left.\frac{d}{d\tilde{p}^2}M^2(p)\right|_{\tilde{p}^2=m^2,p_3=0}&=0,\\
\left.\frac{d^n}{d(p_3^2)^n}M^2(p)\right|_{\tilde{p}^2=m^2,p_3=0}&=0,
\end{align}
where $n=1,2,3,\cdots$. These conditions guarantee that the when the 1PI graphs
are summed to give the full propagator in Eq.~(\ref{propagator}), this 
propagator has the form of a free particle as an asymptotic state. An 
alternative way of expressing this is that the wave function renormalization is
a function of $p_3^2$. If one expands out in powers of $p_3^2$ there will 
be an infinite number of wave function counterterms with one associated with
each term in the expansion. 

Latter we will evaluate the $2\to 2$ scattering amplitude, so we will 
specify the appropriate renormalization condition here. 
The renormalized scattering amplitude is usually defined to take a
certain value at some kinematic point, such as $s=4m^2,t=u=0$, which
then determines the renormalized coupling $\lambda$.
One has the usual
kinematic constraint of the Mandelstam variables $s+t+u=4m^2$ using 
conservation of 4-momentum. 
Again with the loss of Lorentz invariance one 
must impose a renormalization
condition which specifies the renormalized coupling in a scattering
amplitude oriented 
in a certain way with respect to the underlying lattice in space. 
One must also specify the values of the 
3-components of the momentum. There are two values that must be specified.
For scattering $\phi(p_i)\phi(p_i')\to \phi(p_f)\phi(p_f')$, define 
$p^s_3=(p_1+p_2)_3^2$, $p^t_3=(p_1-p_3)_3^2$, and $p^u_3=(p_1-p_4)_3^2$. 
Specification of these quantities for the 
renormalization condition then properly and unambiguously
defines the renormalized coupling.
For the 4-point function on $M^3\times S^1$, we use the following condition  
\begin{align}
\left.iM_4(p)\right|_{s=4m^2, t=u=0, (p_i+p_i')_3=(p_i-p_f)_3=(p_i-p_f')_3=0}
=-i\lambda,
\end{align}
where $s,t,u$ are Mandelstam variables defined in the usual way in terms of 
squared four-momenta.
Using momentum conservation 
the conditions on the 3-components of the momenta is equivalent to the 
condition $p_{i3}=p_{i3}'=p_{f3}=p_{f3}'=0$ (as measured in the preferred 
frame). Since the Lorentz symmetry is broken, a definition of the renormalized
$\lambda$ requires these additional specifications to uniquely 
define the renormalization point.

We calculate one-loop correction of two point function derived by
$\Phi^4$ interaction on $M^3\times S^1$. 
Since the Lorentz symmetry is broken there must appear different wave function
renormalization constants for the $M^3$ and the $S^1$ directions. These can 
be derived in the standard way from the Lagrangian corresponding to the 
Hamiltonian defining our theory.
\begin{align}
-iM^2(p)&=
\frac{-i\lambda}{2}\int_{-\infty}^{+\infty}\frac{d^3k}{(2\pi)^3}
\int_{-\pi/\ell}^{+\pi/\ell}\frac{dk_3}{2\pi}\nonumber\\
&\ \ \ 
\times\frac{i}{k_0^2-k_1^2-k_2^2
-\frac{4}{\ell^2}\sin^2\left(\frac{\ell k_3}{2}\right)-m^2}\nonumber\\
&\ \ \
+i\left(\tilde{p}^2\delta_Z
+\sum_{n=1}^\infty\frac{1}{n!}(p_3^2)^n\delta_{Z_3n}-\delta_m\right),
\end{align}
where $\delta_m$, $\delta_Z$ and $\delta_{Z_3n}$ are counterterms, and 
the tree-level propagator is 
\begin{align}
\frac{i}{p_0^2-p_1^2-p_2^2
-\frac{4}{\ell^2}\sin^2\left(\frac{\ell p_3}{2}\right)-m^2}.
\end{align}
After some calculation, we obtain 
\begin{align}
&-iM^2(p)\nonumber\\
&=\frac{i\lambda m}{4\pi^2\ell}
E\left(-\frac{4}{\ell^2m^2}\right)
+i\left(\tilde{p}^2\delta_Z
+\sum_{n=1}^\infty\frac{1}{n!}(p_3^2)^n\delta_{Z_3n}-\delta_m\right),
\end{align}
where $E(x)$ is the elliptic integral of the second kind, and we applied 
the dimensional regularization scheme and $\overline{\rm MS}$ for
three-dimensional Minkowski spaces. 
The one loop diagram induced from the $\lambda\Phi^4$ term has no 
momentum dependence. For $1/\ell\gg m$, we obtain 
\begin{align}
\delta_Z&=0,\\
\delta_{Z_3n}&=0,\\
\delta_m&=\frac{i\lambda m}{4\pi^2\ell}
E\left(-\frac{4}{\ell^2m^2}\right)\nonumber\\
&=\frac{\lambda}{32\pi^2}
\left[\frac{16}{\ell^2}
+m^2\log\left(\frac{4}{\ell^2m^2}\right)
+m^2\log 2\right]
+O\left(\ell^2\right),
\end{align}
where we used
\begin{align}
E(-x)&=\sqrt{x}+\frac{1}{\sqrt{x}}
\left[-\frac{1}{4}\log\left(\frac{1}{x}\right)+\frac{1}{4}
\log 2\right]+O(1/x^{3/2})
\end{align}
for $x\to\infty$.
There is no wave function renormalization for either the usual continuum case
$M^4$ or the discrete case $M^3\times S^1$ 
at the one-loop level.
An interesting feature of the propagator is that the loop corrections require
a set of wavefunction renormalization constants $\delta_{Z_3n}$. These 
constants are required to produce a renormalized propagator which is 
periodic in $p_3$ and satisfies the renormalization conditions. 
Nevertheless the wavefunction renormalization
factors $\delta_{Z_3n}$ are not independent of each other since they must 
conspire to produce a renormalized propagator defined on $M^3\times S^1$ even
away from its pole.

As already remarked, at the one-loop level there is no momentum dependence 
(and thus no wave function renormalization) for the propagator 
in the $\phi^4$ theory. There is nothing 
interesting to say about its analytic structure. Therefore
let us now proceed to 
calculate four-point function on $M^3\times S^1$. We should
first define the renormalization condition as setting $\lambda$ equal to
the magnitude of the scattering amplitude at zero momentum.
We can write the amplitude as
\begin{align}
iM_4&=-i\lambda+(-i\lambda)^2
\left[iV(s;p_3^s)+iV(t;p_3^t)+iV(u;p_3^u)\right]\nonumber\\
&\ \ \ -i\delta_\lambda,
\end{align}
where
{\small
\begin{align}
&(-i\lambda)^2\cdot iV(p^2;p_3)\nonumber\\
&:=\int\frac{d^4k}{(2\pi)^4}
\frac{(-i\lambda)^2}{2}
\frac{i}{k_0^2-k_1^2-k_2^2-\frac{4}{\ell^2}
\sin^2\left(\frac{\ell k_3}{2}\right)-m^2}\nonumber\\
&\times 
\frac{i}{(k_0+p_0)^2-(k_1+p_1)^2-(k_2+p_2)^2
-\frac{4}{\ell^2}\sin^2\left(\frac{\ell (k_3+p_3)}{2}\right)-m^2}.
\end{align}}

We wish to examine the analytic properties of this amplitude. It is well known
that in the continuum there is a branch cut which represents the dispersive
part of the four-point diagram. It accounts for the imaginary part of the 
amplitude and appears when the value of $p^2$ is sufficient to create real 
(not virtual) particle in the loop. 

Our intent here is to 
take seriously the theory defined on a lattice 
(in one discrete dimension $M^3\times S^1$) and
investigate the branch cut structure of the propagator. Typically lattice 
physicists are interested in a situation where one is close to the continuum 
limit ($p^2 \ll 1/\ell^2$). In the continuum  
a branch point appears at the real particle thresholds, and this branch point
must appear in its usual place in the discretized theory. In the continuum
the branch cut extending away from the branch point goes to infinity. We wish
to demonstrate in this paper 
that in the discretized theory the branch cut ends on 
another branch point. This new branch point arises from the fact that the
momentum space is a circle.
 
After we integrate out three dimensional Minkowski space-time $M^3$ by
using dimensional regularization, we obtain the following integral
\begin{widetext}
\begin{align}
V(\tilde{p}^2,p_3)&=-\frac{1}{32\pi^2}
\int_0^1dx\int_{-\pi/\ell}^{+\pi/\ell}dk_3
\frac{1}{\left[
(1-x)\frac{4}{\ell^2}\sin^2\left(\frac{\ell k_3}{2}\right)
+x\frac{4}{\ell^2}\sin^2\left(\frac{\ell(k_3+p_3)}{2}\right)
+m^2-x(1-x)\tilde{p}^2\right]^{1/2}}\nonumber\\
&=-\frac{1}{32\pi^2}
\int_0^1dx\int_{-\pi/2}^{+\pi/2}dy
\frac{2/\ell}{\left[\frac{4}{\ell^2}
\left\{(1-x)\sin^2y
+x\sin^2(y+z)\right\}
+m^2-x(1-x)\tilde{p}^2\right]^{1/2}},
\label{fourpoint}
\end{align}
\end{widetext}
where $y=\ell k_3/2,z=\ell p_3/2$ and $\tilde{p}^2=p_0^2-p_1^2-p_2^2$.
For illustration of the 
fundamental analytic properties of the propagator we choose
$\tilde{p}^2 \gg 1/\ell^2$ and $p_3=0$. For this subset of cases the theory
effectively has no degrees of freedom in the 3-direction.
In this limit one expects the theory to 
behave as one effectively in $2+1$ dimensions as the lattice spacing is much 
larger than the inverse momenta. The integral in Eq.~(\ref{fourpoint}) becomes
 
\begin{align}
\int_{-\pi/2}^{+\pi/2}dy
\frac{1}{\left(\tilde{\Delta}+\sin^2y\right)^{1/2}}
&=\frac{2}{\sqrt{1+\frac{1}{\tilde{\Delta}}}\sqrt{\tilde{\Delta}}}
K\left(\frac{1}{1+\tilde{\Delta}}\right)
\nonumber\\
&=\frac{2}{\sqrt{\tilde{\Delta}}}
K\left(-\frac{1}{\tilde{\Delta}}\right),
\label{elliptic}
\end{align}
where $K(x)$ is the elliptic integral of the first kind
and the dimensionless quantity
$\tilde{\Delta}=\ell^2 \Delta/4=\ell^2(m^2-x(1-x)\tilde{p}^2)/4$. 

\begin{figure}
        \centering
                \includegraphics[totalheight=2.5in]{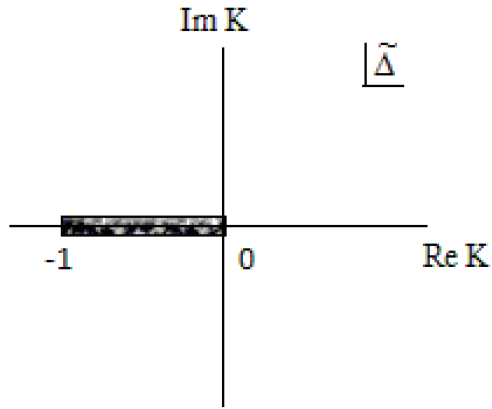}
        \caption{The branch cut of the elliptic integral.}
        \label{fig:branchcut}
\end{figure}

The behavior of the elliptic integral $K(1/\tilde{\Delta})$ in the complex plane
is shown in Fig.~\ref{fig:branchcut}. There are two branch points. The one at 
$\tilde{\Delta}=0$ is the usual branch point associated with the onset of the 
dispersive behavior of the propagator which survives taking the continuum limit.
The other branch point at $\tilde{\Delta}=-1$ corresponds to the onset of 
dispersive behavior in the dimensionally reduced $2+1$ theory where the 
3-direction degrees of freedom are frozen out. It is clear that the branch 
point here (and in more general cases) emerges when the denominator of the 
integrand in Eq.~(\ref{elliptic}) passes through zero. 
The appropriate $i\epsilon$ 
prescription in the propagator (hidden in the treatment above) places one 
it on one side of the branch cut as usual.

It is worthwhile investigating the nature of the branch point that arises
at $\tilde{\Delta}=-1$. 
The branch point at $\tilde{\Delta}=0$ occurs for small loop momentum $k_3$ for
which the integrand in Eq.~(\ref{fourpoint}) is
\begin{align}
&\frac{1}{\left[
\frac{4}{\ell^2}\sin^2\left(\frac{\ell k_3}{2}\right)
+m^2-x(1-x)\tilde{p}^2\right]^{1/2}}\nonumber\\
&\to  \frac{1}{\left[k_3^2
+m^2-x(1-x)\tilde{p}^2\right]^{1/2}},
\end{align}
which is of the same nature of the branch point in the continuum case. In fact
it becomes this branch point when $\ell\to 0$.
If one now examines the integral for a four-point diagram in 
Eq.~(\ref{fourpoint}) (again for the case $p_3=0$) near the point 
$\tilde{\Delta}=-1$, one has
\begin{align}
&\frac{1}{\left[
\frac{4}{\ell^2}\sin^2\left(\frac{\ell k_3}{2}\right)
+m^2-x(1-x)\tilde{p}^2\right]^{1/2}}\nonumber\\
&\to  \frac{1}{\left[-k_3^2
+\left (m^2+\frac{4}{\ell^2}\right )-x(1-x)\tilde{p}^2\right]^{1/2}},
\end{align}
which has the interpretation as a branch cut with momentum-squared $-k_3^2$ 
and an 
effective mass $M^2=m^2+4/\ell^2$.
When the continuum limit $\ell\to 0$ is taken this mass approaches infinity.
The effect of the minus sign is to reorient the direction of the branch cut so 
that it extends to the right, and the $i\epsilon$ prescription places the 
amplitude on one side of the cut. The conclusion here is that if one takes the
field theory description to remain valid for all momentum scales rather than
being an effective theory, then the 
branch cut ends at a new branch point. 

The new branch point that arises is associated with the existence of 
a maximum contribution to the energy from the discretized 
3-direction\cite{Peskin:1995ev}
\begin{align}
E_k^2=k_1^2+k_2^2+\frac{4}{\ell^2}\sin^2\left(\frac{\ell k_3}{2}\right)+m^2,
\end{align}
At this new point the propagator of the 
theory recovers a Lorentz-type symmetry of the same size
as occurs in the continuum limit.

One can reintroduce nonzero values of $p_3$ and perform the integral in 
Eq.~(\ref{fourpoint}). 
An expansion in $\ell p_3$ yields coefficients 
which involve elliptic integrals. The amplitude is presented 
here as an expansion appropriate for small $\ell^2$,
i.e. close to the continuum,
\begin{align}
iM_4&=
-i\lambda-\lambda^2\left[
iV(\tilde{s},p_3^s)+iV(\tilde{t},p_3^t)+iV(\tilde{s},p_3^u)
\right]-i\delta_\lambda,
\end{align}
where $\tilde{s}=(\tilde{p}_i+\tilde{p}^\prime)^2$, etc.  
The counter term is
\begin{align}
\delta_\lambda=-\lambda
\left[V(4m^2,0)-2V(0,0)\right],
\end{align}
and 
\begin{widetext}
\begin{align}
iM_4&=-i\lambda+\frac{i\lambda^2}{32\pi^2}\int_0^1dx
\bigg[
-\log\left(\frac{\Delta_s}{\Delta_{4m^2}}\right)
+\frac{1}{4}
\left\{
\tilde{\Delta}_s
\log\left(\frac{\tilde{\Delta}_{\tilde{s}}}{16}\right)
-\tilde{\Delta}_{4m^2}
\log\left(\frac{\tilde{\Delta}_{4m^2}}{16}\right)
\right\}\nonumber\\
&\ \ \
-x(1-x)z^2\log\left(\frac{\tilde{\Delta}_{\tilde{s}}}{16}\right)
+\frac{1}{2}\left(
\tilde{\Delta}_{s}-\tilde{\Delta}_{4m^2}\right)
\nonumber\\
&\ \ \ -\frac{7}{4}x(1-x)z^2
+\frac{1}{24\tilde{\Delta}_{\tilde{s}}}x(1-x)(8-45x+45x^2)z^4\nonumber\\
&\ \ \ +\frac{1}{24\tilde{\Delta}_{\tilde{s}}^2}x^2(1-x)^2(-8+35x-35x^2)
z^6+O(z^8;\ell^2)
+(s\leftrightarrow t)+ (s\leftrightarrow u)\bigg]\nonumber\\
&=-i\lambda+\frac{i\lambda^2}{32\pi^2}\int_0^1dx
\bigg[
-\log\left(\frac{\Delta_s}{\Delta_{4m^2}}\right)
-\log\left(\frac{\Delta_t}{m^2}\right)
-\log\left(\frac{\Delta_u}{m^2}\right)\nonumber\\
&\ \ \ 
+\frac{\ell^2}{16}
\left\{\Delta_s\log\Delta_{\tilde{s}}+
\Delta_t\log\Delta_{\tilde{t}}+\Delta_u\log\Delta_{\tilde{u}}
-\Delta_{4m^2}\log\Delta_{4m^2}-2m^2\log m^2\right\}\nonumber\\
&\ \ \
-\frac{\ell^2}{4}x(1-x)
\left[
p_3^{s2}\log\left(\frac{\ell^2\Delta_{\tilde{s}}}{64}\right)
+p_3^{t2}\log\left(\frac{\ell^2\Delta_{\tilde{t}}}{64}\right)
+p_3^{u2}\log\left(\frac{\ell^2\Delta_{\tilde{u}}}{64}\right)
\right]
\nonumber\\
&\ \ \ -\frac{7}{16}\ell^2x(1-x)
(p_3^{s2}+p_3^{t2}+p_3^{u2})
+\frac{1}{96}\ell^2x(1-x)(8-45x+45x^2)
\left(\frac{p_3^{s4}}{\Delta_{\tilde{s}}}
+\frac{p_3^{t4}}{\Delta_{\tilde{t}}}
+\frac{p_3^{u4}}{\Delta_{\tilde{u}}}
\right)
\nonumber\\
&\ \ \ +\frac{1}{96}\ell^2x^2(1-x)^2(-8+35x-35x^2)
\left(\frac{p_3^{s6}}{\Delta_{\tilde{s}}^2}
+\frac{p_3^{t6}}{\Delta_{\tilde{t}}^2}
+\frac{p_3^{u6}}{\Delta_{\tilde{u}}^2}\right)
+O(p_3^8;\ell^2)\bigg].
\end{align}
\end{widetext}

One could in principle use this expression to obtain corrections to the 
usual one-loop cross section coming from the nonzero discretization ($\ell$).
Rather than pursue this direction (which, if $\ell$ is taken to be the Planck
length to perhaps model some discretization arising from a theory of 
quantum gravity, gives experimentally small and therefore 
uninteresting results), we comment on
the form of the expression which follows from its analytical properties in the 
complex plane.
The appearance of terms of the form $\ell^2\Delta \log (\Delta)$ are 
indicative of the new branch point that exists at $\Delta =-1$ that 
terminates the usual branch cut in the amplitude that appears in the continuum.
The extension to nonzero values of $p_3$ does not qualitatively alter this 
behavior of the amplitude in the complex plane. 

Notice that the behavior of the scattering 
amplitude in Eq.~(\ref{fourpoint}) arises 
precisely because the momentum integral is performed over a circle.
The finite extent of the branch cut would survive 
the inclusion of higher order quantum corrections because the 
renormalized propagator involves a periodic function of $p_3$. The 
compactification of momentum space introduces another branch point on which 
the usual branch cut evident in the continuum can terminate. 
This structure is also independent of how the renormalization conditions are 
chosen. It should also clear that the existence of two branch points will 
also be the case if one discretizes more of the spatial directions as 
they arise when the denominator in the integral for the scattering amplitude 
passes through zero.

\section{Conclusion}

We have considered the quantum corrections in a theory with one spatial
direction put
on a lattice. This coupling is represented by a periodic 
function in the compactified momentum. This way of viewing the momentum space
is dual to the notion of a compactified configuration space and lends insight 
into the nature of the renormalized propagator. 

The propagator and the four-point diagrams 
in the interacting theory were computed (to one loop). 
The analytic structure of the scattering 
amplitudes involves a branch cut which 
originates in the usual place when considered in terms of the continuum limit.
This branch point is associated with the threshold for real particles as 
required by the optical theorem. 
However, rather than extending to infinity in momentum space, it terminates 
at another branch point which results as a direct consequence of the 
discretization of the spatial direction. 
The mathematical nature of the lattice corrections to the continuum are 
dictated in part by the new branch point. 
A fundamental theory addressing Planck scale physics should account how the 
branch cut extends into regions which involve Planck scale momenta. In the 
toy model we considered here, we can employ field theory at all length scales,
and (at least in this case) the 
branch cut terminates at a another branch point.

\section*{Acknowledgments}

This work is supported in part
by DOE grant DE-FG02-91ER40661.

\end{document}